\documentclass[preprint]{vgtc}               

\ifpdf
  \pdfoutput=1\relax                   
  \usepackage{graphicx}                
  \DeclareGraphicsExtensions{.pdf,.png,.jpg,.jpeg} 
\else
  \usepackage{graphicx}                
  \DeclareGraphicsExtensions{.eps}     
\fi%

\graphicspath{{figures/}{pictures/}{images/}{./}} 

\usepackage{microtype}                 
\PassOptionsToPackage{warn}{textcomp}  
\usepackage{textcomp}                  
\usepackage{mathptmx}                  
\usepackage{times}                     
\usepackage{cite}                      
\usepackage{tabu}                      
\usepackage{booktabs}                  

\onlineid{0}

\vgtccategory{Research}

\vgtcinsertpkg

\title{Designing for Ambiguity: Visual Analytics in Avalanche Forecasting}

\author{Stan Nowak\thanks{e-mail: snowak@sfu.ca}\\ %
        \scriptsize Simon Fraser University %
\and Lyn Bartram\thanks{e-mail: lyn@sfu.ca}\\ %
     \scriptsize Simon Fraser University %
\and Pascal Haegeli\thanks{e-mail: pascal\_haegeli@sfu.ca.}\\ %
     \parbox{1.4in}{\scriptsize \centering Simon Fraser University }}

\abstract{Ambiguity, an information state where multiple interpretations are plausible, is a common challenge in visual analytics (VA) systems. We discuss lessons learned from a case study designing VA tools for Canadian avalanche forecasters. Avalanche forecasting is a complex and collaborative risk-based decision-making and analysis domain, demanding experience and knowledge-based interpretation of human reported and uncertain data. Differences in reporting practices, organizational contexts, and the particularities of individual reports result in a variety of potential interpretations that have to be negotiated as part of the forecaster's sensemaking processes. We describe our preliminary research using glyphs to support sensemaking under ambiguity. Ambiguity is not unique to public avalanche forecasting. There are many other domains where the way data are measured and reported vary in ways not accounted explicitly in the data and require analysts to negotiate multiple potential meanings. We argue that ambiguity is under-served by visualization research and would benefit from more explicit VA support.%
} 

\CCScatlist{
  \CCScatTwelve{Human-centered computing}{Visu\-al\-iza\-tion}{Visu\-al\-iza\-tion application domains}{Visual analytics};
}

\begin{document}


\firstsection{Introduction}

\maketitle

Uncertainty is an important issue in visualization, but to date,  visualization research has primarily focused on uncertainties that are explicit in data and how to represent them \cite{fernandes_uncertainty_2018, greis_uncertainty_2018, hullman_pursuit_2019, hullman_why_2016, kale_hypothetical_2019, thomson_typology_2005, brodlie_review_2012}. However, uncertainty concerns much more than just data, and researchers have highlighted the need to explicitly consider uncertainties that result from reasoning processes in analysis \cite{maceachren_visual_2015, zuk_visualization_2007}. One such critical type of uncertainty that arises in complex and collaborative analyses is \textit{ambiguity}, which we define as an informational state in which multiple interpretations may be equally plausible. It pertains to a multiplicity of potential ways to interpret, rather than simply a lack of or erroneous data. While uncertainty in data itself may lead to ambiguity, it is only one aspect in many that are relevant to consider. While VA researchers have recently focused on more explicit support for knowledge-based and interpretative processes in VA \cite{choi_concept-driven_2019, choi_visual_2019, andrienko_viewing_2018}, ambiguity, to our best knowledge, has not been considered.

We partnered with Avalanche Canada (AC), a public avalanche forecasting organization, to develop VA tools for assessing avalanche hazard. Our collaboration revealed that ambiguity is an important challenge in avalanche forecasting and meaningfully accounting for it is critical to the design of helpful VA. This paper presents the results and lessons learned from our case study. We discuss how challenges in data and analysis give rise to ambiguity, our design strategies to facilitate sensemaking in the face of ambiguity, and feedback from forecasters who have used our tools. The contributions of this paper include an applied discussion of the role and nature of ambiguity in complex analysis and decision-making applications such as avalanche forecasting and a preliminary design trajectory for improved VA support for sensemaking under ambiguity. A key finding from our initial research is that ambiguity cannot---and should not---be "designed away", but rather acknowledged and sometimes encouraged for richer sensemaking. 

\section{Avalanche Forecasting}
Avalanches are natural disaster phenomena where an instability in the snowpack stratigraphy releases a mass of snow that slides downhill with destructive force. They pose a significant risk to those working and recreating in mountainous terrain. Avalanche forecasting is concerned with predicting current and future snow instabilities that may result in avalanches through human or natural triggers \cite{mcclung_elements_2002}.  Forecasters produce daily avalanche hazard assessments that communicate avalanche hazard conditions and risks to those traveling through avalanche-prone mountainous terrain. As with the forecasting of floods, wildfires, hurricanes, and other similar extreme weather events, avalanche forecasting involves risk prediction and communication where the audience varies in role and expertise, from the general public to those experienced with extreme natural hazards.

 Avalanche forecasting is viewed as a largely inductive or Bayesian-like process where new information is used to continuously update mental models of avalanche conditions formed over the course of an entire season \cite{mcclung_elements_2002-1}. Forecasters assess and characterize avalanche hazards by answering a series of questions: (1) \textit{what} types of avalanches exist? (2) \textit{Where} are they located? (3) \textit{How} likely are avalanches to occur? (4) \textit{How} large will they be? \cite{statham_conceptual_2018}. To answer these questions, Forecasters utilize a variety of observations and data sources as part of the avalanche hazard assessment process \cite{mcclung_elements_2002-1}. 
 
 While many forecasters have the benefit of physically working in the field within small-scale regions producing assessments for an expert audience, AC public avalanche forecasters assess hazards and communicate them to the public from an office-based setting, relying heavily on second-hand reports, and applying them to large regions that often experience significant variability in avalanche conditions. Such reports are provided by the Canadian Avalanche Association's Industry Information Exchange (InfoEx) \cite{haegeli_infoex_2014}. It is a web-based platform where professional avalanche safety operations such as ski resorts, helicopter skiing operations, or operations dealing with transportation corridors such as railways or highways share local observations and assessments. Reported data include field observations of weather, snow conditions, avalanche activity, descriptions of locations traveled to and hazard assessments formalized by industry standards. Data are viewed in large text tables with minimal visualization support. 
 
 The complexity of avalanche phenomena and the variability of operational contexts from which these data are sourced leave forecasters to rely heavily on subjective judgement and knowledge to discern context and fill gaps in understanding. This process leaves room for multiple potential interpretations and ambiguities. The challenges of interpretation under complexity and uncertainty are similar to those faced in the forecasting of many other types of natural disasters \cite{beven_epistemic_2018} as well as risk management applications that utilize human-produced data \cite{diehl_user-uncertainty:_2018}.

\section{Defining Ambiguity}
We distinguish ambiguity as a separate issue from that of data uncertainty. Ambiguity has been described as dealing with the reliability, credibility, or adequacy of risk information \cite{ellsberg_risk_1961}, as well as a multiplicity of states \cite{smithson_ignorance_1993} or outcomes \cite{ayyub_elicitation_2001} among many others. We define it as a multiplicity of plausible interpretations. This definition is more closely related to the philosophical or semiotic notions of ambiguity \cite{sennet_ambiguity_2016}.

Ambiguity emerges out of sensemaking under complexity: comprehensive understanding of a complex system is intractable given limitations of human perception and observation \cite{grabowski_simple_2008}. These limitations mean that prediction involves more subjective judgement, speculation, and imagination than it does precise deductions (this is often referred to as mental simulation \cite{hancock_role_2018}). Sensemaking involves drawing on prior knowledge to negotiate alternative interpretations of the problem at hand \cite{klein_data-frame_2007}. In this way, sensemaking is more about resolving multiple meanings, rather than simply accounting for missing information \cite{klein_data-frame_2007}. Recent research has highlighted how data scientists intervene and transform data based on their intuitive and knowledge-based understanding to better support more meaningful sensemaking processes \cite{muller_how_2019}. Choices made in the process of analysis, however defensible, can lead to widely different analytic results, even when following high standards of scientific rigor \cite{silberzahn_many_2018}. Ambiguity is ubiquitous in complex analysis because sensemaking involves much more than data itself. 

When sensemaking is shared, ambiguity arises not only from the varying perspectives of the collaborators, but also the complexities of communication \cite{heer_design_2008}. In large organizations or settings where analysis is shared across a variety of operational contexts, ambiguity is common because as soon as analysis is shared it loses its context \cite{prue_overcoming_2014}. The relevance of any piece of information is context-sensitive as information itself is defined by the relations between data, the world it represents, and observers' goals, expectations, and interests \cite{woods_diagnosis_2002}. Further, sensemaking processes involve prior knowledge that may not be explicitly visible in shared data or information \cite{pirolli_sensemaking_2005}, leaving collaborators to make inferences based on their own knowledge and experiences. Hence, domains that rely heavily on analyzing data and assessments produced by others such as public health \cite{mccurdy_framework_2019}, intelligence analysis \cite{prue_overcoming_2014}, search and rescue operations, crisis management, avalanche forecasting, and many others are likely to experience challenges of ambiguity.

\section{Case Study}
In collaboration with the AC development team we designed a set of VA tools to support daily operational avalanche hazard assessment and address it's accompanying challenges of sensemaking under ambiguity. The forecasters prioritized two types of observations that are essential in avalanche hazard assessment: weather observations from weather stations which provide a source of ground truth for meteorological forecast validation, and structured field reports of avalanche observations which are considered to be key indicators characterizing the nature of avalanche hazards \cite{mcclung_elements_2002-1}.

\subsection{Study Design}

To meaningfully inform our design process, we initially conducted a series of studies using a suite of qualitative interview and observational methods improving our understanding of the application domain. Through consultations with the avalanche forecasters, we iteratively developed a set of prototypes. Think-aloud studies and unstructured interviews were used to solicit feedback for improving the designs. We used historical and synthetic data to evaluate the avalanche observation tool, while the weather tool used real-time data and was used operationally during the 2019/2020 winter season. At the end of the winter season, we conducted semi-structured interviews to allow forecasters to reflect on how the prototypes addressed the challenges of avalanche forecasting and how they changed the nature of work. Eight avalanche forecasters participated in the interviews. Seven were actively involved throughout the design process. One forecaster simply provided feedback based on their experience using the weather tool. Forecasters were interviewed using a video conferencing tool. These interviews were recorded, transcribed, and qualitatively coded for common themes that emerged.

\subsection{Weather Observations}
Public avalanche forecasters use telemetry readings from remote weather stations as a form of ground truth to validate previous forecasts and to track weather systems throughout the workday.

\subsubsection{Data and Analysis Challenges}
 The spatial distribution of these weather stations is sparse relative to the spatial heterogeneity of weather patterns that these weather stations are expected to measure, challenging the reliability of such meteorological observations \cite{lundquist_our_2019}. To account for such variability, forecasters focus on familiar weather stations and use their experiences and knowledge of how local terrain at weather station sites interacts with weather systems to inform interpretation of such data.

\subsubsection{Design}

Traditionally, forecasters have had to view station data individually through a variety of web portals. Our visualizations provide aggregate descriptive statistics summarizing recent spatio-temporal patterns in precipitation, wind, and temperature. They also show individual station data through tooltips.

\subsubsection{Forecaster Feedback}

 Forecasters found it challenging to adjust their traditionally bottom-up approach to fit with the top-down overview our tool provided, yet nevertheless found the tool to enrich their sensemaking process. While our design approach was not especially novel, forecasters feedback revealed how forecasters used the tool to support their inductive and Bayesian-like reasoning processes. They noted that numerical aggregations do provide a useful overview and starting point, but that the impressions that such an overview leaves have to be refined using an iterative process that involves imagining how weather station telemetry at individual stations translate from its local geographic context into broader regional patterns. This process is as much speculative, involving the formation and evaluation of various plausible expectations, as it is observational, grounding these expectations in data that captures traces of the material world. This informed our subsequent designs.

\subsection{Avalanche Observations}
The visualization tool shown in Figure 1 was developed to support the investigation of field reported data characterizing observed avalanches.

\begin{figure*}[ht]
\centering
\includegraphics[width=.88\textwidth]{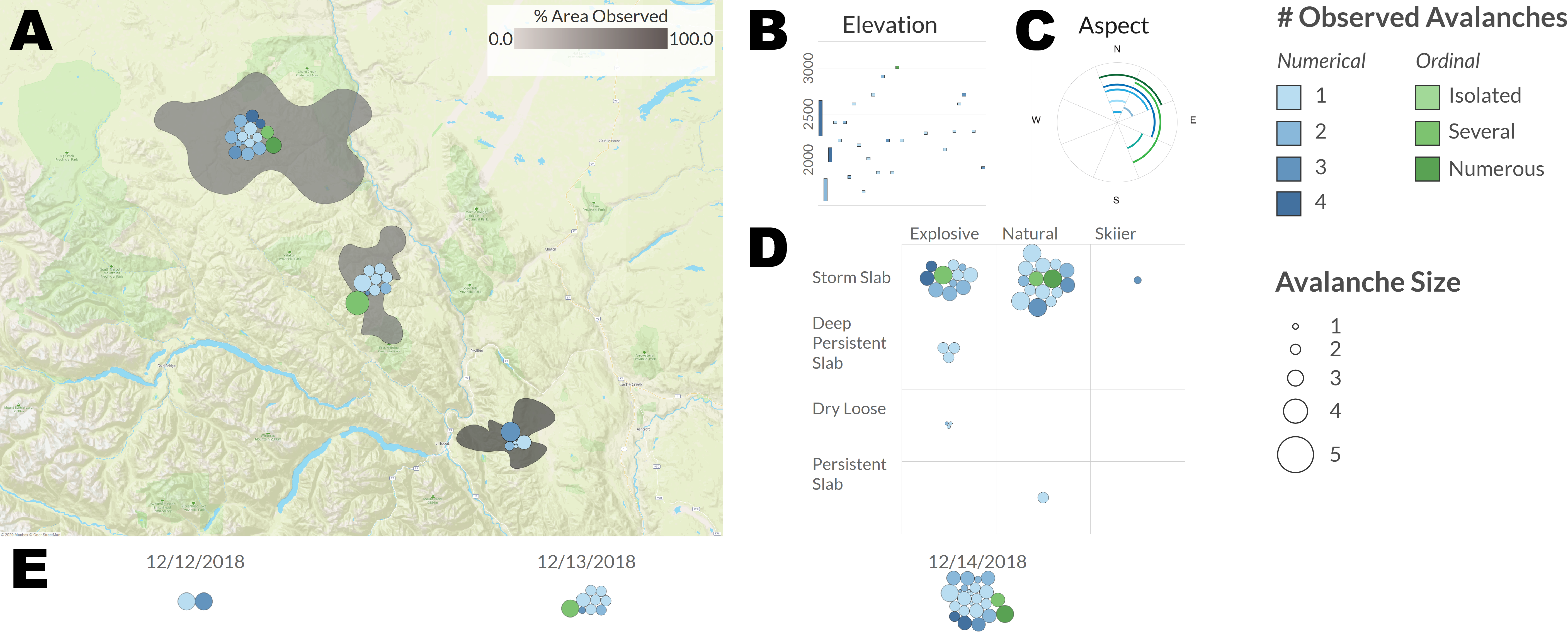}
\caption{Multiple coordinated views featuring unit visualizations and glyphs. The spatial data used in this visualization are synthesized in part to protect the identity of the avalanche safety operations whose data we acquired to develop this tool. Three days-worth of avalanche observation reports are shown. A) Map displaying glyphs positioned at centroids of polygons related to particular avalanche safety operations. B) Chart showing elevation ranges (y-axis) associated with avalanche reports (x-axis). C) Arc diagram showing ranges of cardinal directions (polar axis) associated with avalanche reports (radial axis). D) Grid matrix showing avalanche observations across types of avalanche problems (rows) as well as avalanche trigger types (columns). E) Timeline showing avalanche observations by day. }
\label{fig1}

\end{figure*}

\subsubsection{Data and Analysis Challenges}
Observations of avalanches are the most highly valued information forecasters have at their disposal because they are direct evidence of the existence of an avalanche problem. In the InfoEx, Avalanche observations are described using unstructured text fields and standardized structured fields following the Observation Guidelines and Recording Standards (OGRS) set by the Canadian Avalanche Association \cite{canadian_avalanche_association_observation_1995}. They are used to answer (in part) the essential questions in avalanche hazard assessment. These data explicitly include the \textit{types} of avalanches observed, wherein the \textit{terrain} they were observed (e.g. geo-position, elevation and, "aspect": the compass direction that a mountain slope faces), and what the respective \textit{sizes} of observed avalanche were. However, to understand how \textit{likely} avalanches are requires a more nuanced reading. \textit{Likelihood} is determined through a combination of \textit{sensitivity} to triggering and \textit{spatial distribution} which are inferred from the integration of a wide variety of data attributes such as trigger types, the number of avalanches triggered, geographic position, descriptions of terrain as well as other data and knowledge not included in our visualization tool. This is a deeply interpretative process that leaves forecasters with a holistic understanding of avalanche conditions which is continuously refined with new information.

In spite of standardization, even structured data may be interpreted in a variety of ways due to differences in operational needs, goals, and constraints of the reporting organizations.
 
“\textit{...the InfoEx system and the standards… they kind of define the box that we all work in. But what bits and pieces reside inside that box and which ones you use and how you use them... context drives that… you might... use a certain approach…  data… that are… obviously within that... general framework or box that we've created, but you might not use them exactly the same way.}”\textbf{P8}
 
One example of this is an attribute describing the number of avalanches that were observed. This single attribute can be reported using numerical data or ordinal bins that describe specific ranges for the number of observed avalanches (e.g. "several" is defined as 2-9 avalanches) \cite{canadian_avalanche_association_observation_1995}. Transforming such data into a single and unified data type to enable numerical aggregations and more parsimonious visualizations may appear like a desirable choice given the technical definition. Doing so, however, would hide critical contextual information that forecasters use in their sensemaking process. The choice of whether to use a number or an ordinal bin can indicate much more than the sum of observed avalanches when considering the operational context in which the report was produced. For instance, the use of an ordinal bin can be interpreted as conveying the rate at which avalanches are occurring relative to the amount of terrain that has been observed, as expressing uncertainty around the number of avalanches truly observed, or simply reflect a time-saving grouping of avalanche observations that are characteristically different and would otherwise be reported separately.

This is but one example illustrating the challenges of ambiguity embodied in these data. There are many other fields where forecasters similarly rely on contextual information and the particularities of reports to support meaningful interpretation. As a result, our visualizations needed to improve on sensemaking support of existing tools (text tables) without occluding contextual information. Our success with the weather visualization suggested that an overview would be beneficial, but we could not rely on numerical aggregations. This is not least because the ambiguity of avalanche observation data involves a re-conceptualization of the dimensions or measurements in question, whereas weather station telemetry remain largely conceptually self-consistent.

\subsubsection{Design}

To address the ambiguity related challenges of this data we took a glyph-based approach. Glyphs can operate at multiple scales of resolution and can provide an overview without having to rely on numerical aggregations. Instead, they allow visual aggregation operations such as summarizing data, detecting outliers, detecting trends, or segmenting data into clusters \cite{szafir_four_2016}, while at the same time showcasing granular data to reveal its particularities.

We used a packed bubble chart glyph (shown in Figure 1A, 1D, 1E). Each circle represents an individual report. The size of reported avalanches is encoded as the size of each circle.   Additionally, colour hue is used to distinguish whether a numerical (blue) or ordinal (green) data type was used to report the number of observed avalanches. Colour saturation/luminance is used to encode value in the number of avalanches observed with darker colours encoding higher values. Circles are organized using a packed layout. The resulting glyph supports various forms of visual aggregation providing a holistic multidimensional overview of the data. For instance, when comparing any two glyphs, the overall size of the glyph, the number of circles, and their sizes and colours, all combine to provide a nuanced and multidimensional view that allows forecasters to evaluate multiple perspectives of the data. 

The glyphs are positioned within a timeline presenting the day avalanches were estimated to have occurred (Figure 1E) and a matrix presenting information about the types of avalanches observed and what triggered them (figure 1D). Glyphs are also positioned at centroids of polygons representing the tenures of the reporting operation (figure 1A). Polygons are shaded according to the estimated percentage of the tenure that was observed.

Elevations and aspects of avalanches are reported using non-uniform intervals. To maintain visibility of each individual report and prevent overplotting, we developed several charts that encode reports as line or arc segments with their lengths representing these non-uniform intervals. Figure 1B shows elevation ranges of reported avalanches with elevation on the y-axis and an index for each individual report on the x-axis. Similarly, Figure 1C shows aspect ranges of reported avalanches with the polar axis encoding aspect and the radial axis serving as an index for each individual report.

Reports are presented across multiple coordinated displays that support standard brushing and highlighting interactions. Selecting reports highlights them in all corresponding visualizations providing a multidimensional perspective into the data. Additionally, tooltips provide access to unstructured data associated with individual reports.

It is important to note that the visual overview provided by such glyphs deliberately uses perceptually weak visual encodings to provide forecasters with a holistic and multidimensional perspective as a starting point for their assessments. Our design is aimed to facilitate the Bayesian-like reasoning processes of avalanche forecasting. As forecasters explore the data, access contextual information, and negotiate meaning in these ambiguous data, they update their mental model of avalanche conditions without being hampered by overly precise visualizations that could prescribe a certain perspective.

Forecasters use this interface to answer the essential questions in avalanche hazard assessment pertaining to avalanche type, likelihood, size, and location in terrain. While this interface supports several low-level visual tasks such as comparison and trend detection to help answer these questions, its primary purpose is to allow forecasters to view and access the particularities of individual reports and mentally tune their understanding of any identified visual patterns. This functionality is non-trivial. Being able to discern what is meant by a particular datum such as the choice to use an ordinal bin for the number of avalanches observed can mean the difference between being able to or not being able to detect central characteristics of avalanche hazards. In this example, the operational context in which the number of avalanches was reported in can influence how to interpret the spatial density of avalanche occurrences in turn influencing the forecaster's perception of likelihood and the entire nature of the avalanche hazard.

We found that this approach is well suited to the task of negotiating the potential meanings of such ambiguous data and analyses. This may not be the only or even the most optimal solution, however, it serves to highlight how sensemaking around ambiguous human reported data can be supported through careful and ambiguity-aware design choices.

\subsubsection{Forecaster Feedback}

Forecasters informed us that our prototype was a good representation of the mental operations they perform using conventional tools. They reflected that our approach is more methodologically sound than purely quantitative approaches.
 
“\textit{I like seeing the individual events more than that aggregates… it seems full of flaws and limitations to kind of summarize all the activity with one number.}” \textbf{P7}

Our prototype was used as a point of reference in internal discussions that surfaced discrepancies in how forecasters interpret certain data, illustrating the potential for visualizations such as these to enrich organizational knowledge and practices.

\section{Discussion}

Field reported data used by AC and the Canadian avalanche industry embody the central challenges of ambiguity in distributed collaborative analysis and decision-making. Standards can provide a common language, but complexity and differences in context create challenges in interpretation that cannot simply be designed away. These data may be reported along seemingly uniform and coherent dimensions, but in context of analysis, they have to be re-conceptualized to consider multiple alternative interpretative lenses at the same time. Our glyph-based visualizations were designed to capture this multidimensional correspondence with meaning, while simultaneously displaying the particularities of each report to allow forecasters to glean enough context to decide what is relevant and how to interpret the visual aggregates that the glyphs provide.

A key take-away from this case study is that identifying and characterizing sources of ambiguity is critical for designing visualizations that are intended to meaningfully support sensemaking. Another is that while there is a seemingly widespread belief that ambiguity is something to always be prevented or reduced, it often serves a critically functional role in sensemaking and is therefore better embraced and specifically designed \textit{for}. We recognize that ambiguity is much broader and arises from many more sources than just human reported data. Our design approach was more about enabling ambiguities to be recognized and reasoned through, rather than explicitly encoded. Our intent is to provide a starting point for future research in this area and encourage researchers to explore challenges of ambiguity in similar complex and shared decision-making applications.

A key challenge that remains is how to address ambiguity more explicitly. When analysis is shared and the same data are revisited as part of subsequent analyses, any identified ambiguities, relevant materials, or knowledge are often not explictly captured in the data. Researchers studying how data workers cope with uncertainty found that when faced with ambiguity, common coping approaches include annotations and references to clarifying materials \cite{boukhelifa_how_2017}. Others have used annotations with templated structured fields to aid public health experts in externalizing their knowledge of measurement errors implicit in international infectious disease reporting \cite{mccurdy_framework_2019}. Supporting awareness of uncertainty is critical in analysis and visualization \cite{sacha_role_2016}. We argue this is also the case for ambiguity and that conventional visualization techniques such as annotations can be extended and specifically tailored to address the challenges of collaborative sensemaking under ambiguity. However, we recognize that externalizing knowledge requires effort and can be disruptive \cite{heer_design_2008}. This is especially the case in constrained and risk-based decision-making applications such as avalanche forecasting. Hence, tailoring existing visualization techniques to provide low-cost mechanisms to address ambiguity introduces a rich research and design space to explore.

\section{Conclusion}
Through a case study with public avalanche forecasters, we discovered that ambiguity, a state in which multiple interpretations are plausible, bears direct relevance to the design of VA systems. In this domain, complexity, data sparsity, and the challenges of communication in distributed collaborations produce ambiguity and leave forecasters to use their knowledge and experience to negotiate interpretation. Our visualization tools addressed ambiguity using a glyph-based approach that enables forecasters to navigate multiple potential interpretation for meaningful sensemaking. While this study highlights the challenges of ambiguity, we argue that research into a more targeted and explicit approach for capturing and representing ambiguities is warranted.

\acknowledgments{
 Thanks to Avalanche Canada, the Vancouver Institute for Visual Analytics (VIVA), the Big Data Initiative at Simon Fraser University (SFU), the SFU Avalanche Research Program, and our reviewers for their thoughtful feedback. This work was supported in part by the Natural Sciences and Engineering Research Council Industry Research Chair in Avalanche Risk Management (grant no. IRC/515532-2016), which receives industry support from Canadian Pacific Railway, HeliCat Canada, Canadian Avalanche Association, and Mike Wiegele Helicopter Skiing.}

\bibliographystyle{abbrv}

\bibliography{References}
\end{document}